\documentclass[final,3p,times]{elsarticle}
\usepackage{graphicx}
\usepackage{amssymb}
\usepackage{a msmath}
\usepackage{hyperref}

\journal{Communications in Nonlinear Science and Numerical Simulation}

\usepackage[usenames]{color}
\usepackage[normalem]{ulem}
 
\begin{document}

\begin{frontmatter}
\title{Phase-separated  vortex-lattice  in a rotating binary  Bose-Einstein condensate}

\author[ift]{ S. K. Adhikari}

\ead{sk.adhikari@unesp.br}

\address[ift]{
Instituto de F\'{\i}sica Te\'orica, UNESP - Universidade Estadual Paulista, \\ 01.140-070 S\~ao Paulo, S\~ao Paulo, Brazil
} 

\begin{abstract}

We study circularly-symmetric phase separation of vortex lattices in a rapidly rotating harmonically-trapped
quasi-two-dimensional    binary Bose-Einstein condensate (BEC)   by introducing a weak quartic trap in one of the components.  The increase of the  rotational frequency 
in such a  system is also found to   generate  a phase separation of the vortex lattices
of an overlapping non-rotating BEC.  The phase-separated vortex lattices have  different structures for a  binary BEC with inter-species repulsion and inter-species attraction. In the former case of a fully repulsive binary BEC, the phase separation of the vortex-lattices is accompanied by a complete phase separation of component densities. In the latter case of   inter-species attraction, there is a partial  phase separation of component densities, although   there could be  a complete phase separation of the generated vortex lattices in the two components. In the case of inter-species attraction, we need to have different intra-species repulsion in the two components for an efficient phase separation. 
We compare and contrast our results with the phase separation obtained in a harmonically-trapped binary BEC without any quartic trap.

\end{abstract}


\begin{keyword}
Rotating Bose-Einstein condensate; Gross-Pitaevskii equation; Split-step Crank-Nicolson scheme;
 Vortex lattice
\end{keyword}

\end{frontmatter}

\section{Introduction}

 Soon after the observation  of a trapped Bose-Einstein condensate (BEC) at ultra-low density and temperature in a laboratory \cite{becexpt,becexpt2}, rotating 
trapped condensates  were created and studied. Under slow rotation a small number of vortices were created \cite{vors}. With the increase of rotational frequency $\Omega$, large vortex arrays 
were generated \cite{vorl}. The vortices have quantized circulation as in super-fluid $^4$He in bulk: \cite{Sonin,fetter}
$\oint_{\cal C} {\bf v} .  d{\bf r}={2\pi\hbar n}/{m},$
where ${\bf v}({\bf r},t)$ is the super-fluid velocity field, ${\cal C}$ is a generic closed path, 
$m$ is the mass of an atom 
and $n$ is an integer denoting the quantized angular momentum of an atom in units of $\hbar$ in the trapped rotating BEC.
As the rotational frequency increases in a rotating BEC,    it is energetically favorable to form a lattice of quantum vortices of unit circulation each ($n=1$) \cite{fetter}. Consequently, a rapidly rotating trapped BEC is found to form a large number of vortices of unit circulation each arranged with a definite symmetry often in a Abrikosov  triangular lattice \cite{vorl,abri}.    
Because of the weak-coupling low-density limit of the trapped BEC, it has been possible to study the formation of  vortices in it by the mean-field   Gross-Pitaevskii (GP) equation.

With the advance of experimental techniques,  vortex lattice structure with a large number of vortices has been observed \cite{Schweikhard,Hall} and studied theoretically \cite{wang,thlatsep} in a binary BEC. There has also been study of vortex-lattice formation in a BEC along the weak-coupling to unitarity crossover \cite{sci}.
 The study of vortex lattices in a binary or a multi-component spinor BEC is interesting because the interplay between inter-species and intra-species interactions may lead to the formation of  square  vortex  lattice \cite{Schweikhard,kumar}, other than the standard Abrikosov triangular lattice \cite{abri}.    In addition there could be  vortices of fractional charge \cite{frac,Cipriani}, coreless vortices \cite{coreless} and phase-separated vortex lattices in multi-component spinor  \cite{thlatsep} and dipolar  \cite{kumar}   BECs. The difficulty of 
experimental study of overlapping vortices in  different components of  a multi-component or a binary BEC is monumental and despite great interest in the study of vortex lattices in a binary BEC \cite{Kasarev2,Mueller,Barnett,Wei,Kuopanportti,Kasashet}, this has limited the number of these  studies \cite{Schweikhard,Hall} . Hence for experimental investigation it will be useful to have phase-separated vortex lattices in a binary BEC, where the vortices of one component do not overlap with the vortices of the other component.

In a repulsive homogeneous BEC,  phase separation takes place for  \cite{phse}
\begin{equation} \label{eq1}
\frac{g_1g_2}{g_{12}^2}<1, 
\end{equation}
where $g_1$ and $g_2$ are intra-species repulsion strengths for components 1 and 2, respectively, and $g_{12}$ inter-species repulsion strength.  This useful condition, although not rigorously valid in a trapped quasi-two-dimensional (quasi-2D)  system, may
provide an approximate guideline for phase separation. 
It is well-known  that a phase separated vortex lattice in a binary BEC can be generated  by manipulating the parameters $g_1,g_2,$ and $g_{12}$ \cite{thlatsep,phse}.

  In this paper we consider a mechanism for efficient generation of circularly-symmetric 
phase-separated vortex lattices  in a harmonically trapped quasi-2D  binary BEC
 by including  a weak quartic trap \cite{quartic} in  the first component when the first component stays on an inner circle and the second on a concentric ring, viz. Figs. \ref{fig1}(c)-(d).
  Circularly-symmetric states with the symmetry  axis coinciding with the axis of rotation can efficiently generate a large number of phase-separated vortices arranged with a definite symmetry. 
Unless the parameters $g_1$ and $g_2$ are very different and $g_1g_2/g_{12}$ of Eq. (\ref{eq1}) is much less than one,
 the natural phase separation of a harmonically-trapped  binary BEC results in a separation of two components in semicircles in opposite directions breaking the rotational circular symmetry
 \cite{thlatsep} and such a state does not  develop a good lattice structure under rotation.  
The weak  quartic trap can cause an efficient  phase separation in cases, where a phase separation would be impossible without the quartic trap, e. g. for $g_1=g_2$ and/or $g_1g_2/g_{12} \approx 1  $. 
Moreover, rapid rotation  is found to enhance the phase separation of vortex lattices of a binary BEC with a weak quartic trap in one component.
We find that  if we subject such a  non-rotating binary BEC with overlapping components   to rapid rotation,    phase-separated 
 vortex lattices may result.  No such phase separation in an overlapping binary BEC
is possible under rapid rotation in the absence of the weak quartic trap.

 Using mean-field GP equation 
we illustrate the phase separation of vortex lattices  (a) in a harmonically-trapped binary BEC in the presence of a weak quartic trap in one of the components  and (b)  under rapid rotation of  the same with overlapping components.  In addition to considering the phase separation of vortex lattices for repulsive inter-species interaction, we extend our study to the case of attractive inter-species interaction \cite{referee}.  In the case of inter-species attraction, 
there is a partial overlap between the two components; but vortex lattice could be  generated in the phase-separated region of the component(s). In both cases  $-$ inter-species repulsion and attraction  $-$ we contrast the present phase separation of vortex lattices with that obtained in the  absence of the weak quartic trap in one component.

In Sec. II  the mean-field model for a   rapidly rotating binary BEC is presented.  Under a tight trap in the transverse direction a quasi-2D version of the model is also given, which we use in the present study. 
The results of numerical calculation are shown in Sec. III.  
Finally, in Sec. IV we present a brief summary of our findings.

\section{Mean-field model for a rapidly rotating binary BEC}

We consider a binary rotating BEC  interacting via  inter- and intra-species 
interactions.  
The angular frequencies for the axially-symmetric harmonic trap  
along $x$, $y$ and $z$ directions are taken as 
$\omega_x=\omega_y=\omega$ and 
$\omega_z=\lambda\omega$. In addition a quartic trap in the $x-y$ plane will sometimes be  considered in the first component \cite{quartic}. { Now it is possible to subject the two components of a binary BEC 
to different confining traps \cite{ming}, when the two components are two distinct atomic species or two  nearby atomic isotopes, such as rubidium isotopes $^{87}$Rb and $^{85}$Rb \cite{becexpt}, or caesium isotopes $^{133}$Cs, $^{135}$Cs, and $^{137}$Cs \cite{cs} used in BEC experiments.  In such cases of near-by atomic masses, the masses of the two species can be taken to be equal for all practical purpose. Thus in this theoretical study we will take the masses of two species to be equal. }
 The quartic trap under rotation leads 
to a phase separation of the vortex lattice of the two components. 

The study of a rapidly rotating binary BEC is conveniently performed in the rotating frame, where the generated vortex  
lattice is a stationary state \cite{fetter}, which can be obtained by the imaginary-time propagation method \cite{imag}. 
Such a dynamical equation in the rotating frame can be written if we note that the Hamiltonian in the rotating frame is given by  $H = H_0-\Omega l_z$, where $H_0$ is that in the laboratory frame, $\Omega $  is the angular frequency of rotation,  $l_z$ is the $z$ component of angular momentum given by $l_z= i\hbar (y\partial/\partial  x - x \partial/\partial y )$ \cite{ll1960}.
However if the rotational frequency $\Omega$ is increased beyond the trapping frequency $\omega$ the rotating bosonic gas makes a quantum phase transition to a non-superfluid state, where the validity of a mean-field description of the rotating bosonic gas is questionable \cite{fetter}.    
With the inclusion of the extra rotational energy  $-\Omega l_z$ in the Hamiltonian,   the coupled GP
equations for the binary  BEC in the rotating frame for $\Omega <\omega $  can be written as \cite{quartic}
\begin{align} \,
{\mbox i} \hbar \frac{\partial \phi_1({\bf r},t)}{\partial t} &=
{\Big [}  -\frac{\hbar^2}{2m}\nabla^2 -\Omega l_z \nonumber
+ \frac{1}{2}m \omega^2 
(\rho^2+\kappa \frac{m\omega}{\hbar}\rho^4+\lambda^2{z}^2 )+ \frac{4\pi \hbar^2}{m}{a}_1 N_1 \vert \phi_1({\bf r},t)\vert^2
\nonumber
\\  &  
+\frac{4\pi \hbar^2}{m} {a}_{12} N_2 \vert \phi_2({\bf r},t)|^2
{\Big ]}  \phi_1({\bf r},t),
\label{eq1x}
\\  
\label{eq2}
{\mbox i} \hbar \frac{\partial \phi_2({\bf r},t)}{\partial t} &=
{\Big [}  -\frac{\hbar^2}{2m}\nabla^2+ \frac{1}{2}m \omega^2 
(\rho^2+\lambda^2{z}^2 )-\Omega l_z
\nonumber\\ &
+ \frac{4\pi \hbar^2}{m}\Big\{ {a}_2 N_2 \vert \phi_2({\bf r},t) \vert^2
+ {a}_{12} N_1 \vert \phi_1({\bf r},t) \vert^2\Big\}
\Big] 
 \phi_2({\bf r},t),
\end{align}
where   the two species of atoms of mass $m$ each are denoted $i=1,2$, $\phi_i({\bf r},t)$ are the order parameters of the two components, 
$N_i$ is the number of atoms in species 
$i$,  $\quad {\mbox i}=\sqrt{-1}$, ${\bf r}= \{x,y,z\},$  $ {\pmb \rho}=\{ x,y\}$, $\rho^2=x^2+y^2$, $a_i$ is the intra-species scattering length of species $i$,  $a_{12}$ is the inter-species scattering length, 
  $\kappa$ is the strength of the quartic trap included in the first component. The functions $\phi_i$ are normalized as $\int d{\bf r}|\phi_i({\bf r},t)|^2 =1.$

The following   dimensionless form of Eqs. (\ref{eq1x}) and (\ref{eq2})  can be obtained  by  the  transformation of variables: ${\bf r}' = {\bf r}/l_0, l_0\equiv \sqrt{\hbar/m\omega}$, $t'=t\omega,  \phi_i'=   \phi_i l_0^{3/2},  \Omega'=\Omega/\omega, l_z '= l_z/\hbar$ etc.:   
\begin{align}& \,
{\mbox i} \frac{\partial \phi_1({\bf r},t)}{\partial t}=
{\Big [}  -\frac{\nabla^2}{2 }
+ \frac{1 }{2} (\rho^2+\kappa \rho^4+\lambda^2 z^2 ) -\Omega l_z
+ 4\pi N_1a_1 \vert \phi_1 \vert^2 + 4\pi a_{12}N_2\vert \phi_2 \vert^2
{\Big ]}  \phi_1({\bf r},t),
\label{eq3}\\
& \,
{\mbox i} \frac{\partial \phi_2({\bf r},t)}{\partial t}={\Big [}  
- \frac{\nabla^2}{2}+ \frac{1 }{2} (\rho^2+\lambda^2 z^2 ) -\Omega l_z
+ 4\pi N_2 a_2 \vert \phi_2 \vert^2 
+ 4\pi a_{12} N_1 \vert \phi_1 \vert^2 
{\Big ]}  \phi_2({\bf r},t),
\label{eq4}
\end{align} 
where for simplicity we have dropped the prime from the transformed variables.

For a quasi-2D binary BEC in the $x-y$ plane under a strong trap along the $z$ 
direction ($\lambda \gg 1$), the essential vortex dynamics will be confined to  the $x-y$ plane with the $z$ dependence playing a passive role.   The wave functions  can then be written as 
$\phi_i({\bf r},t)= \psi_i({\pmb \rho},t)\Phi(z)$, where the function $ \psi_i({\pmb \rho},t)$ carries the essential vortex dynamics and $\Phi(z)$ is a normalizable Gaussian function. In this case the 
$z$ dependence can be integrated out \cite{luca} and we have the following 2D equations
\begin{align}& \,
{\mbox i} \frac{\partial \psi_1({\pmb \rho},t)}{\partial t}=
{\Big [}  -\frac{\nabla^2}{2 }
+ \frac{1 }{2} (\rho^2+\kappa \rho^4) -\Omega l_z
+ g_1 \vert \psi_1 \vert^2 + g_{12} \vert \psi_2 \vert^2
{\Big ]}  \psi_1({\pmb \rho},t),
\label{eq5}\\
& \,
{\mbox i} \frac{\partial \psi_2({\pmb \rho},t)}{\partial t}={\Big [}  
- \frac{\nabla^2}{2}+ \frac{1 }{2} \rho^2 -\Omega l_z 
+ g_2 \vert \psi_2 \vert^2 
+ g_{21} \vert \psi_1 \vert^2 
{\Big ]}  \psi_2({\pmb \rho},t),
\label{eq6}
\end{align}
where
$g_1=4\pi a_1 N_1\sqrt{\lambda/2\pi},$
$g_2= 4\pi a_2 N_2\sqrt{\lambda/2\pi}  ,$
$g_{12}={4\pi } a_{12} N_2 \sqrt{\lambda/2\pi} ,$
$g_{21}={4\pi } a_{12} N_1  \sqrt{\lambda/2\pi}.$ In this study we will take $N_1=N_2$ which will make $g_{12}=g_{21}$ maintaining the possibility $g_1\ne g_2$ and consider $\Omega<1$ \cite{fetter}.

\section{Numerical Results}

\begin{figure}[!t]

\begin{center}
\includegraphics[trim = 0cm 0.0cm 0cm 0mm, clip,width=.24\linewidth]{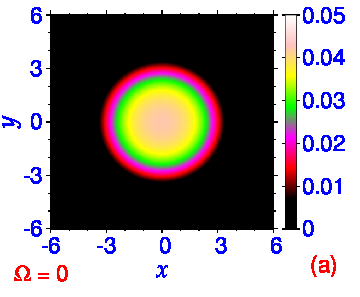}
\includegraphics[trim = 0cm 0.0cm 0cm  0mm, clip,width=.24\linewidth]{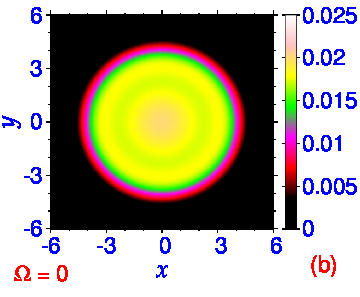}
\includegraphics[trim = 0cm 0.0cm 0cm 0mm, clip,width=.24\linewidth]{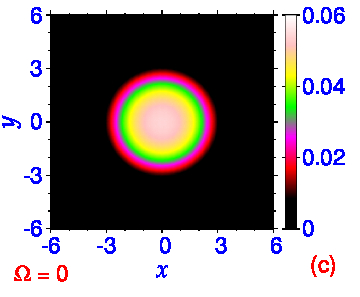}
\includegraphics[trim = 0cm 0.0cm 0cm 0mm, clip,width=.24\linewidth]{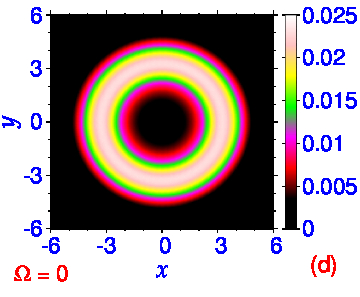}

\includegraphics[trim = 0cm .0cm 0cm 0cm, clip,width=.24\linewidth]{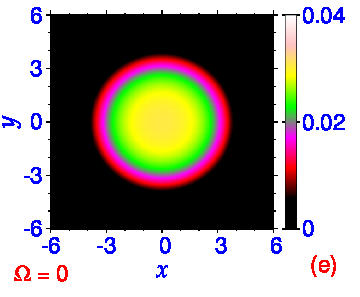}
\includegraphics[trim = 0cm 0cm 0cm 0cm, clip,width=.24\linewidth]{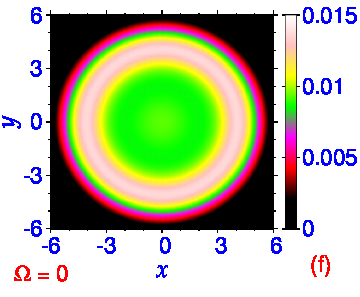}
\caption{   Phase separation from quartic potential ($\kappa =0.05 $)    in non-rotating ($\Omega =0$) binary BEC from a contour plot of 2D densities ($|\psi_i|^2$): (a) first and (b) second components for $g_{12}=150, g_1=g_2=250$, (c) first and (d) second components for 
$g_{12}=228, g_1=g_2=250 $ (e) first and (f) second components for $g_{12}=400$,  $g_1=g_2=600 $. All quantities plotted in this and following figures are dimensionless.
}
\label{fig1}
\end{center}

\end{figure}

The quasi-2D binary mean-field equations
 (\ref{eq5}) and (\ref{eq6}) cannot be solved analytically and different numerical methods, such as the split time-step Cranck-Nicolson method \cite{imag,CPC} or the pseudo-spectral method \cite{PS}, can be employed  for their solution. Here we solve  Eqs. (\ref{eq5}) and (\ref{eq6}) by the split time-step
Crank-Nicolson discretization scheme using a space step of 0.05
and a time step of 0.0002.  There are different
C and FORTRAN programs for solving the GP equation \cite{imag,CPC}  and one should use the appropriate one.
These programs have recently been adapted to simulate the vortex lattice in a rapidly rotating BEC \cite{cpckk} and we use these in this study. 
 In this paper, without considering a specific atom, we will present the results in dimensionless units for different sets of  parameters: 
$\Omega, g_1, g_2, g_{12} (=g_{21}), \kappa$. 
In the phenomenology of a specific atom, the parameters   $g_1, g_2, g_{12}$ can be varied experimentally through a variation of the underlying intra- and inter-species scattering 
lengths 
 by the Feshbach resonance technique \cite{fesh}.

First we consider the symmetric ($g_1=g_2$)  and fully repulsive ($g_i,g_{12}>0$) non-rotating BEC ($\Omega=0$).   We demonstrate a phase separation in this case in the presence of a weak quartic trap in the first component for an inter-species non-linearity $g_{12}$ above a critical value.
 Throughout  this study we will take $\kappa=0.05$ in all examples of weak quartic trap. To illustrate the phase separation in a non-rotating binary BEC, for an example, we  consider $g_1=g_2=250$ and vary $g_{12}$. 
In Figs. \ref{fig1}(a) and (b) we show the contour plot of 2D density of the two components ($=|\psi_i|^2$) 
for $g_1=g_2=250, g_{12}=150, \kappa=0.05$. We find that the first component shown in (a)  due to a tiny quartic trap has smaller size than the second component shown in (b), but there is no phase separation.   
 In Figs. \ref{fig1}(c) and (d) we show the same for $ g_{12}=228$, where there is a circularly-symmetric phase separation due to the increased repulsion between the components with the increase of $g_{12}$: the small-sized  first component mostly occupy a hollow on the central part of the second component.  In Figs. \ref{fig1}(c)-(d) $g_1=g_2$ and $g_1g_2/g_{12}=1.2$ violating condition (\ref{eq1}). A harmonically-trapped binary BEC with the parameters of Figs. \ref{fig1}(c)-(d)  will lead to completely overlapping states,  demonstrating that the 
phase separation
in Figs. \ref{fig1}(c)-(d) is caused by the weak quartic trap.
 In the absence of the quartic trap and for $g_{12}$ above a critical value, the ground state is a phase separated circularly-asymmetric state  with two components separating in semicircular shapes  \cite{thlatsep}. A circularly-symmetric phase separation is possible in a harmonically-trapped binary BEC  for  $g_1$  very different from $g_2$  and  the parameter $g_1g_2/g_{12}$  much smaller than one.  
 In Figs. \ref{fig1}(e) and (f)
we display the contour plot of 2D density for $g_1=g_2= 600, g_{12}=400, \kappa=0.05$. In this case there is no phase separation between the components, although the sizes of the two components are larger due to increased non-linearities $g_i, g_{12}$ compared to those shown in Figs.  \ref{fig1}(a) and (b).  There will be a phase separation in   Figs. \ref{fig1}(e)-(f) if a larger value of $g_{12}$ is chosen as in   Figs. \ref{fig1}(c)-(d) .

\begin{figure}[!t]

\begin{center}
\includegraphics[trim = 0cm 0.64cm 0cm 0mm, clip,width=.24\linewidth]{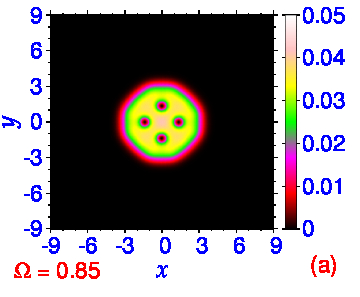}
\includegraphics[trim = 0cm 0.64cm 0cm 0mm, clip,width=.24\linewidth]{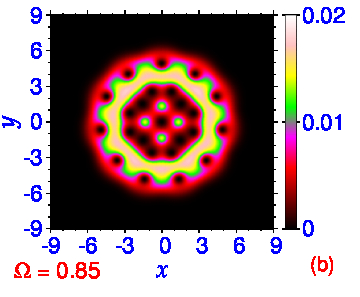}
\includegraphics[trim = 0cm 0.64cm 0cm 0mm, clip,width=.24\linewidth]{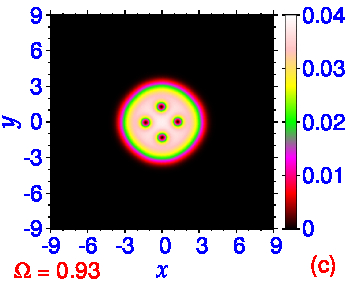}
\includegraphics[trim = 0cm 0.64cm 0cm 0cm, clip,width=.24\linewidth]{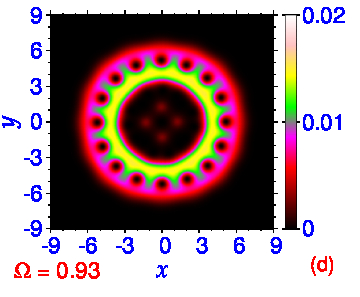}

\includegraphics[trim = 0cm 0.64cm 0cm 0mm, clip,width=.24\linewidth]{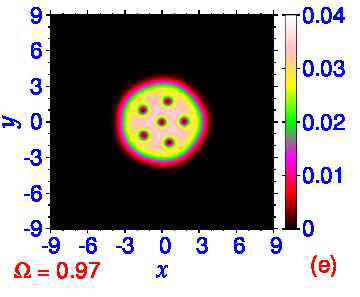}
\includegraphics[trim = 0cm 0.64cm 0cm 0cm, clip,width=.24\linewidth]{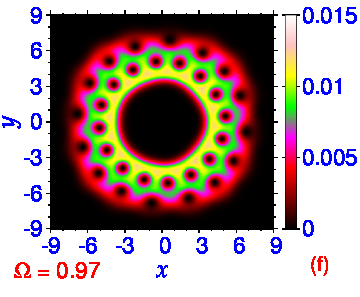}
 
\caption{   Phase-separated vortex lattices in a rapidly rotating binary BEC  from a contour plot of 2D densities ($|\psi_i|^2$): (a) first and (b) second components for $\Omega=0.85$, (c) first and (d) second components for $\Omega=0.93$,   (e) first and (f) second components for $\Omega=0.97$. Other parameters are 
 $g_1=g_2=250, g_{12}=g_{21}=150, \kappa =0.05,$ corresponding to the non-rotating BEC of Fig. \ref{fig1}(a)-(b).
}
\label{fig2}
\end{center}

\end{figure}

\begin{figure}[!t]

\begin{center}
\includegraphics[trim = 0cm 0.64cm 0cm 0mm, clip,width=.24\linewidth]{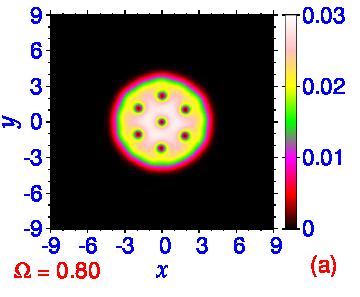}
\includegraphics[trim = 0cm 0.64cm 0cm 0mm, clip,width=.24\linewidth]{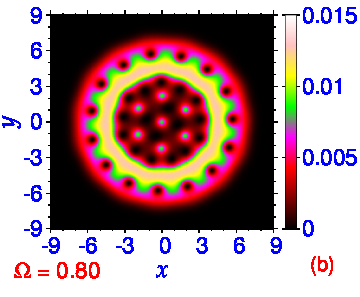}
\includegraphics[trim = 0cm 0.64cm 0cm 0mm, clip,width=.24\linewidth]{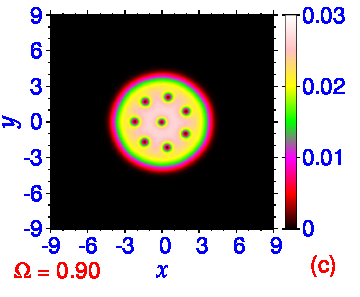}
\includegraphics[trim = 0cm 0.64cm 0cm 0cm, clip,width=.24\linewidth]{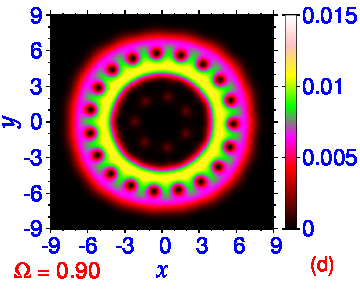}
\includegraphics[trim = 0cm 0.64cm 0cm 0mm, clip,width=.24\linewidth]{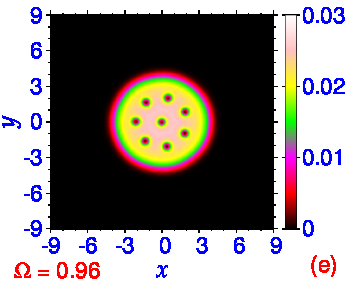}
\includegraphics[trim = 0cm 0.64cm 0cm 0cm, clip,width=.24\linewidth]{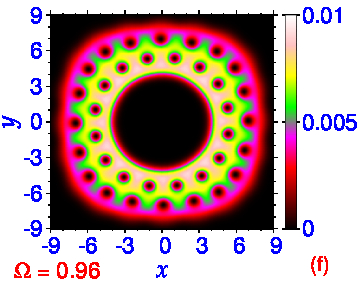}
 
\caption{  Same as in Fig. \ref{fig2}: (a) first and (b) second components for $\Omega=0.80$, (c) first and (d) second components for $\Omega=0.90$,   (e) first and (f) second components for $\Omega=0.96$.  Other parameters are 
 $g_1=g_2=600, g_{12}=g_{21}=400, \kappa =0.05$,
corresponding to the non-rotating BEC of Fig. \ref{fig1}(e)-(f).
}
\label{fig3}
\end{center}

\end{figure}

\begin{figure}[!t]

\begin{center}
\includegraphics[trim = 0cm 0.64cm 0cm 0mm, clip,width=.24\linewidth]{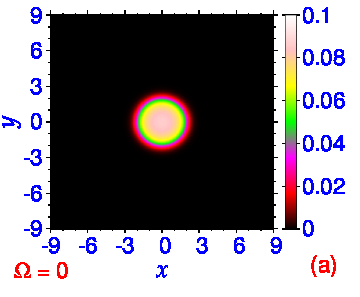}
\includegraphics[trim = 0cm 0.64cm 0cm 0mm, clip,width=.24\linewidth]{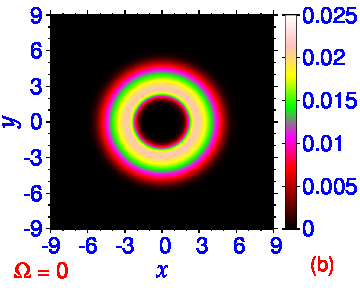}
\includegraphics[trim = 0cm 0.64cm 0cm 0mm, clip,width=.24\linewidth]{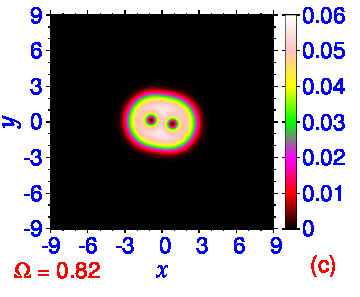}
\includegraphics[trim = 0cm 0.64cm 0cm 0cm, clip,width=.24\linewidth]{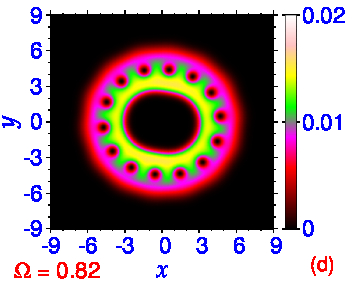}
\includegraphics[trim = 0cm 0.64cm 0cm 0cm, clip,width=.24\linewidth]{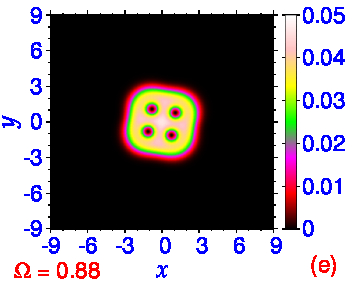}
\includegraphics[trim = 0cm 0.64cm 0cm 0cm, clip,width=.24\linewidth]{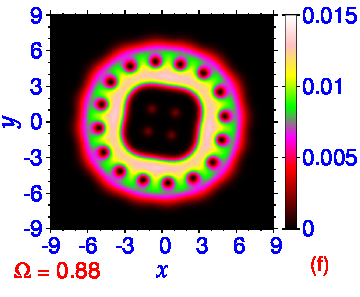}
\includegraphics[trim = 0cm 0.64cm 0cm 0cm, clip,width=.24\linewidth]{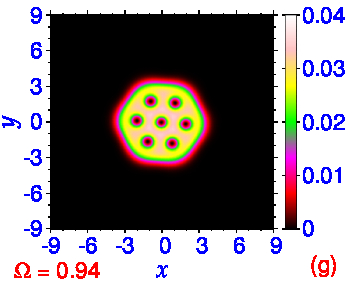}
\includegraphics[trim = 0cm 0.64cm 0cm 0cm, clip,width=.24\linewidth]{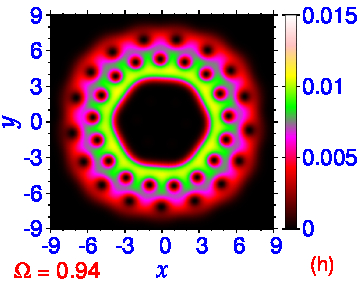}
\caption{   Same as in Fig. \ref{fig2}: (a) first and (b) second components for $\Omega=0$ (non-rotating), (c) first and (d) second components for $\Omega=0.82$, (e) first and (f) second components for $\Omega=0.88$, 
(g) first and (h) second components for $\Omega=0.94$. Other parameters are 
 $g_1=100, g_2=500, g_{12}=g_{21}=300, \kappa =0$.
}
\label{fig4}
\end{center}

\end{figure}

    If we consider the binary BEC exhibited in Figs. \ref{fig1}(c)-(d), where there is a phase separation in the absence of rotation ($\Omega=0$), and subject it to rapid rotation, quite expectedly,  we will have 
phase-separated vortex lattice in the two components.   However, quite unexpectedly,  we find that in the case of the binary 
BECs shown in  Figs. \ref{fig1}(a)-(b) and in  Figs. \ref{fig1}(e)-(f) where there is no phase separation in the absence of rotation, phase-separated vortex lattice appears if subjected to rapid rotation. Rapid rotation acting in conjunction with a tiny quartic trap in one of the components
facilitates a phase separation of the generated vortex lattice.  It is well known that, in the  absence of the  weak quartic trap, rapid rotation does not generate phase separation in a binary BEC. To demonstrate this we subject the binary BEC of Figs. \ref{fig1}(a)-(b) to rapid rotation. The result of imaginary-time simulation of Eqs. (\ref{eq5}) and (\ref{eq6}) 
for non-zero $\Omega$ obtained with the parameters of Figs. \ref{fig1}(a)-(b)  are presented in 
Figs. \ref{fig2}(a)-(f) where we show the contour plot of density of the two components for $\Omega=0.85, 0.93$ and 0.97.  For $\Omega=0.85$ the phase separation is partial with a significant amount of atoms occupying the central region of the second component, viz. Fig. \ref{fig2}(b).   With the increase of $\Omega$, for $\Omega =0.93$, we find that the phase separation is almost complete, illustrated by the quasi-black central region of the second component,   viz. Fig. \ref{fig2}(d). For $\Omega =0.97$ the phase separation is complete as seen by the black central region  in Fig. \ref{fig2}(f). In the first component, under a weak quartic trap superposed on a harmonic trap, the generated vortex lattice has the usual triangular form. In the second  component, the vortices appear in circular arrays around the central hole.  The vortices on different circular arrays are arranged in a triangular  lattice, viz. Fig. \ref{fig2}(f).

 The evolution of the binary BEC of Figs. \ref{fig1}(e)-(f) upon rotation is shown in Figs. 
\ref{fig3}(a)-(f), where we illustrate the contour density plots of the two components for $\Omega= 0.80, 0.90,$  and 0.96. For $\Omega=0.80$ the phase separation is partial,  for 
 $\Omega=0.90$ it is almost complete, and for   $\Omega=0.96$ it is complete.   The numbers of vortices in Figs. \ref{fig2}(a)-(f) for $\Omega =0.85,0.93,$ and 0.97, respectively, are smaller than those 
in the corresponding plots Figs. \ref{fig3}(a)-(f) for smaller  $\Omega =0.80,0.90,$ and 0.96
because the non-linearities in  the latter are larger than those in the former. 
The larger non-linearities in Fig. \ref{fig3} have generated a larger number of vortices for slightly smaller values of rotational frequencies $\Omega$.

The usual way of generating a phase-separated circularly-symmetric  harmonically-trapped non-rotating binary BEC ($\kappa=0$) in the ground state  is through an imbalance in the repulsive intra-species interaction  ($g_1$ very different from  $g_2$) in the presence of inter-species repulsion consistent with the approximate guideline (\ref{eq1}).  
 This is illustrated in Figs. \ref{fig4}(a)-(b) through a contour plot of 2D densities of a non-rotating binary BEC for 
$g_1=100, g_2=500, g_{12}=300, \Omega =0, \kappa=0$. For $g_1=g_2$ such a phase separated ground state does not possess circular symmetry. 
Once this binary BEC is subject to a rapid rotation,  the phase separation continues and vortices appear with the increase of angular frequency of rotation $\Omega$. This is illustrated in Fig. \ref{fig4}(c)-(h) for $\Omega =0.82,
0.88,0.94$ through a contour plot of component densities.  
 In the first component with smaller non-linearity,  the generated vortex lattice has hexagonal structure. In the second component the vortices are generated in circular arrays around the central  hollow: the vortices in different circular arrays tend to stay in a triangular lattice.  The coupling between  the two components reduces the number of vortices in both components.   
For example, if the inter-species coupling between the two components is removed and we consider a binary BEC with $g_1=100, g_2=500, g_{12}=0, \Omega =0.94$ the number of generated vortices in the first and second components are 19 and 43, respectively, whereas the numbers of vortices in the presence of inter-species coupling in Figs.   \ref{fig4}(g)-(h) are 7 and 32, respectively, due to an increase in effective non-linearity in presence of inter-species coupling. If we had considered a smaller inter-species non-linearity, for example  $g_{12}=200$ in place of 300 in Figs. \ref{fig4}(a)-(b) maintaining the  intra-species non-linearities unchanged ($g_1=100, g_2=500$) the phase separation will disappear for $\Omega=0$. No phase
separation will result  under rapid rotation in this case as shown in Figs. \ref{fig2} and \ref{fig3}. 
The presence of a weak quartic trap in one of the components is essential for generating the phase separation under rapid rotation.

    Next we consider the possibility of phase separation of vortex lattice in the presence of attractive inter-species interaction, e.g. negative $g_{12}$ \cite{referee}.  In this case there cannot be a complete phase separation of the component densities like in the case of repulsive  inter-species interaction.  However, there can be a partial or full phase separation of the generated vortices. We see in the following the conditions for achieving this.
In the symmetric case, for $g_1=g_2$,  there is never a complete phase separation of the vortices even in the presence of a weak quartic trap. For an efficient phase separation of the vortex     
 lattices in the presence of a weak quartic trap we have to take $g_1\ne g_2$.   Such a clean phase separation is not possible in the absence of a quartic trap.

\begin{figure}[!t]

\begin{center}
\includegraphics[trim = 0cm 0.64cm 0cm 0mm, clip,width=.24\linewidth]{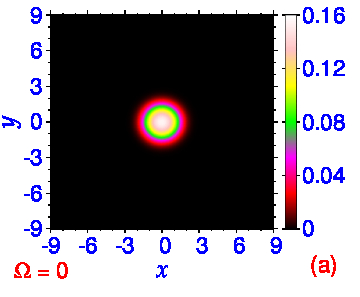}
\includegraphics[trim = 0cm 0.64cm 0cm 0mm, clip,width=.24\linewidth]{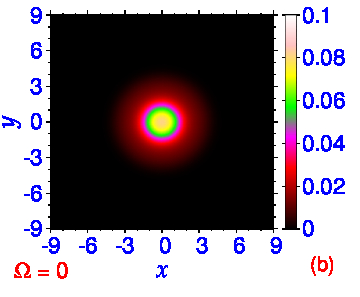}
\includegraphics[trim = 0cm 0.64cm 0cm 0mm, clip,width=.24\linewidth]{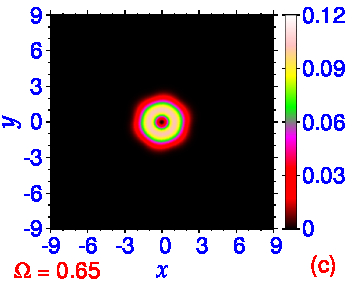}
\includegraphics[trim = 0cm 0.64cm 0cm 0cm, clip,width=.24\linewidth]{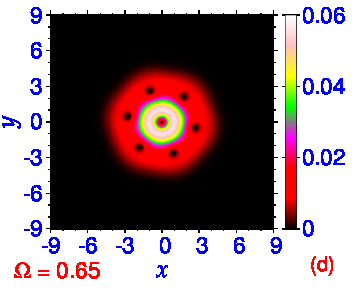}
\includegraphics[trim = 0cm 0.64cm 0cm 0cm, clip,width=.24\linewidth]{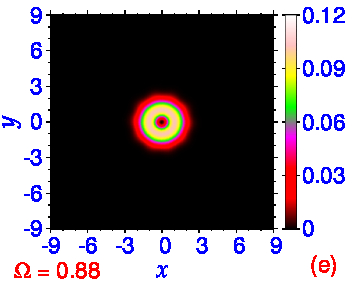}
\includegraphics[trim = 0cm 0.64cm 0cm 0cm, clip,width=.24\linewidth]{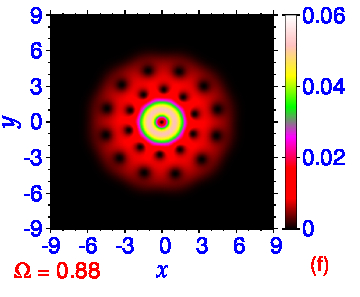}
 \includegraphics[trim = 0cm 0.64cm 0cm 0cm, clip,width=.24\linewidth]{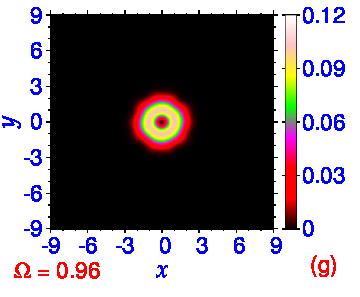}
\includegraphics[trim = 0cm 0.64cm 0cm 0cm, clip,width=.24\linewidth]{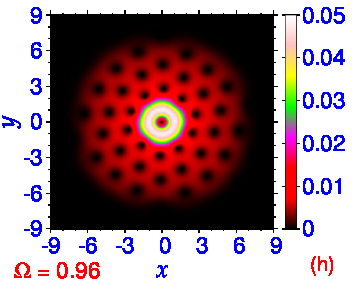}
\caption{  Same as in Fig. \ref{fig2}: (a) first and (b) second components for $\Omega=0$ (non-rotating),  (c) first and (d) second components for $\Omega=0.65$, (e) first and (f) second components for $\Omega=0.88$, (g) first and (h) second components for $\Omega=0.96$. Other parameters are  
 $g_1=100, g_2=500, g_{12}=g_{21}=-200, \kappa =0.05$.
}
\label{fig5}
\end{center}

\end{figure}

\begin{figure}[!t]

\begin{center}
\includegraphics[trim = 0cm 0.64cm 0cm 0mm, clip,width=.24\linewidth]{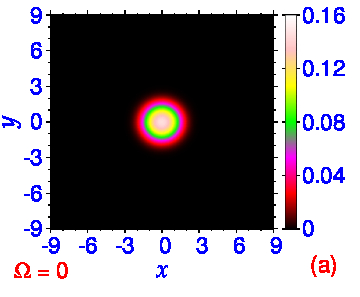}
\includegraphics[trim = 0cm 0.64cm 0cm 0mm, clip,width=.24\linewidth]{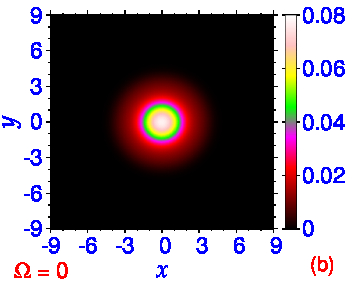}
\includegraphics[trim = 0cm 0.64cm 0cm 0mm, clip,width=.24\linewidth]{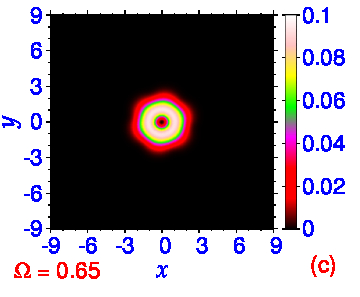}
\includegraphics[trim = 0cm 0.64cm 0cm 0cm, clip,width=.24\linewidth]{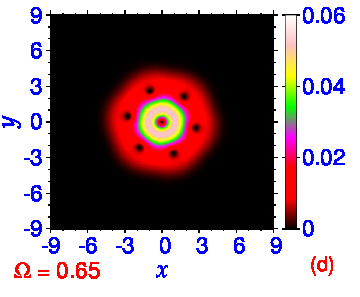}
\includegraphics[trim = 0cm 0.64cm 0cm 0cm, clip,width=.24\linewidth]{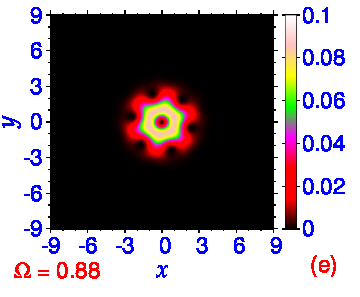}
\includegraphics[trim = 0cm 0.64cm 0cm 0cm, clip,width=.24\linewidth]{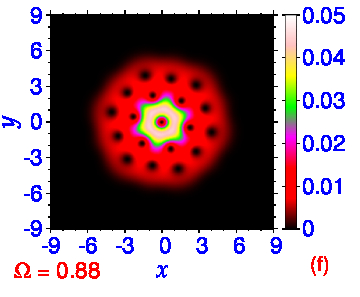}
 \includegraphics[trim = 0cm 0.64cm 0cm 0cm, clip,width=.24\linewidth]{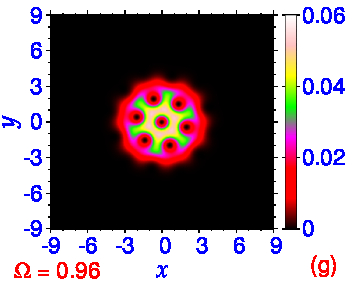}
\includegraphics[trim = 0cm 0.64cm 0cm 0cm, clip,width=.24\linewidth]{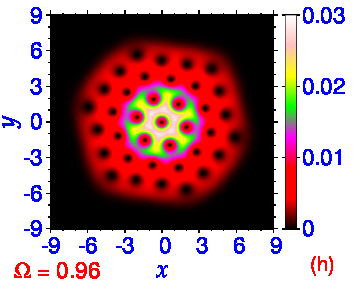}
\caption{  Same as in Fig. \ref{fig5} but with  $\kappa =0$.
}
\label{fig6}
\end{center}

\end{figure}

To demonstrate the phase separation for inter-species attraction, first we consider a binary BEC with $g_1=100, g_2 = 500, g_{12}=-200, \kappa =0.05$ in the presence of a weak quartic trap.
In this case the vortex lattice is generated, under rotation, only in the second component with larger non-linearity.   In Fig. \ref{fig5} we display the contour plots of density of the two components for $\Omega =0, 0.65, 0.88$ and 0.96. In this case the vortices are formed in the second component in closed arrays.  The vortices on different arrays stay on a triangular lattice. Next we consider the vortex lattice generation in the same binary BEC as in Fig. \ref{fig5} without the weak quartic trap ($\kappa=0$). In this case the generation of vortex 
lattices is illustrated in 
Fig. \ref{fig6} for the same rotational frequencies, $\Omega=0, 0.65, 0.88, 0.96$,  as in Fig. \ref{fig5}. For $\kappa=0$,  with the increase of $\Omega$ a vortex lattice appears in the first component  also, viz. Fig. \ref{fig6}, e.g. seven vortices arranged in a hexagonal lattice in Fig. \ref{fig6} (g) for $\Omega = 0.96$. In the second component the vortices are arranged in  perfect hexagonal lattices with 7 vortices in Fig. \ref{fig6}(d), 19  vortices in Fig. \ref{fig6}(f), and 
37  vortices in Fig. \ref{fig6}(h) for $\Omega = 0.65, 0.88$ and 0.96, respectively. There are two differences in generated vortex lattices in Figs. \ref{fig5} and \ref{fig6} with and without the quartic trap. First,  no clean hexagonal lattice is generated in the presence of the quartic trap in Fig. \ref{fig5}.  Secondly, as the quartic trap highly inhibits vortex generation \cite{quartic}, no vortex lattice is generated in the first component in Fig. \ref{fig5}.

\section{Summary and Discussion} 
 
We have studied the generation of  phase-separated circularly-symmetric vortex lattices in a harmonically-trapped 
repulsive quasi-2D  binary BEC with identical intra-species repulsion in the two components and 
with a weak quartic trap included in one of the components.  We find that the weak quartic trap 
facilitates a phase separation of the vortex lattices of the two components for parameter domain leading to overlapping binary BEC in the absence of the quartic trap. 
For example, the phase-separated binary BEC with a weak quartic trap  of Fig. \ref{fig1}(c)-(d) 
becomes overlapping upon removal of the quartic trap. According to guideline (\ref{fig1}) the 
parameters of the binary BEC of Fig. \ref{fig1}(c)-(d) should be overlapping.
 Rapid rotation of a binary BEC with a quartic trap in one component further enhances the phase separation. 
For example, the non-rotating overlapping binary BECs shown in Figs. \ref{fig1}(a)-(b) and in 
Figs. \ref{fig1}(e)-(f) generate  phase-separated vortex lattices under rapid rotation as shown in Figs. \ref{fig2}
and \ref{fig3}, respectively.

We also considered  phase-separated vortex lattices in a  binary BEC with inter-species attraction. In this case there is never a full separation of component densities. An efficient 
separation of the vortex lattices of the two components is possible if we take different intra-species repulsion in the two components. In the presence of the weak quartic trap in one of the components,  it was possible to generate vortex lattice in one of the components in a region where the density of the other component is zero, viz. Fig. \ref{fig5}. In this  case  the vortex lattice is generated in closed arrays with the vortices on adjacent arrays lying on a triangular lattice. 
In the absence of the quartic trap, with the increase of angular frequency $\Omega$, vortices were generated in both components and only a partial separation of vortex lattices of the two components was possible as shown in Fig. \ref{fig6}.  Different from Fig. \ref{fig5}, 
the generated vortex lattice in this case has perfect hexagonal form. With present experimental know-how phase-separated vortex lattices can be generated in a laboratory.

\section*{Acknowledgements}
\noindent

SKA thanks the Funda\c c\~ao de Amparo \`a Pesquisa do
Estado de S\~ao Paulo (Brazil) (Project: 
2016/01343-7) and the Conselho Nacional de Desenvolvimento Cient\'ifico e Tecnol\'ogico (Brazil) (Project:
303280/2014-0) for partial support.


%

\begin{thebibliography}{99}

\bibitem{becexpt}    M. H. Anderson, J. R. Ensher, M. R. Matthews, C. E. Wieman, 
and E. A. Cornell, Science { 269} (1995) 198.


\bibitem{becexpt2} 
K. B. Davis, M. -O. Mewes, M. R. Andrews, N. J. van Druten, D. S. Durfee, D. M. Kurn, and W. Ketterle,
Phys. Rev. Lett. { 75} (1995) 3969.

\bibitem{vors}K. W. Madison,  F. Chevy, W. Wohlleben, and J. Dalibard,
 Phys. Rev. Lett.
{ 84} (2000) 806;  


M. R. Matthews,  B. P. Anderson, P. C. Haljan, D. S. Hall,
M. J. Holland, J. E. Williams, C. E. Wieman, and E. A.
Cornell,  Phys. Rev. Lett. { 83} (1999) 3358.

\bibitem{vorl}J. R. Abo-Shaeer,  C. Raman, J. M. Vogels, and W. Ketterle,
 Science
{ 292} (2001) 476; 

P. C. Haljan,  B. P. Anderson, I. Coddington, and E. A. Cornell,  Phys. Rev. Lett.
{ 86} (2001) 2922.


 \bibitem{Sonin}
E. B. Sonin, 
\textit{Dynamics of Quantised Vortices in Superfluids} 
(Cambridge University Press, Cambridge, 2016).


\bibitem{fetter}  A. L. Fetter, Rev. Mod. Phys. { 81} (2009) 647. 



\bibitem{abri}A. A. Abrikosov,  Zh. Eksp. Teor. Fiz.
{ 32} (1957) 1442; [Eng. Transla.  Sov. Phys.-JETP
{ 5} (1957) 1174.]

\bibitem{Schweikhard}
V. Schweikhard, I. Coddington, P. Engels, S. Tung, and E. A. Cornell, 
Phys. Rev. Lett. {93} (2004) 210403. 

\bibitem{Hall}
D. S. Hall, M. R. Matthews, J. R. Ensher, C. E. Wieman, and E. A. Cornell, 
Phys. Rev. Lett. {81} (1998) 1539. 


 \bibitem{wang} H. Wang, 
J. Sci. Comput. { 38} (2009) 149.
 

\bibitem{thlatsep} P. Mason and A. Aftalion, 
Phys. Rev. A {84} (2011) 033611.

  

\bibitem{sci}S. K. Adhikari and L. Salasnich, Scientific Rep. 8 (2018) 8825.



\bibitem{kumar}
R. K. Kumar, L. Tomio, B. A. Malomed, and A. Gammal, 
Phys. Rev. A {96} (2017) 063624; 

N. Ghazanfari, A. Kele\c{s}, M. \"{O}. Oktel, 
Phys. Rev. A {89} (2014) 025601. 




\bibitem{frac}S.-W. Su,
C.-H. Hsueh,
I.-K. Liu,
T.-L. Horng,
Y.-C. Tsai,
S.-C. Gou,
and W. M. Liu, Phys. Rev.  A { 84} (2011) 023601.

\bibitem{Cipriani}
M. Cipriani and M. Nitta, 
Phys. Rev. Lett. {111} (2013) 170401.

\bibitem{coreless}A. E. Leanhardt, Y. Shin, D. Kielpinski, D. E. Pritchard, and W. Ketterle,
Phys. Rev. Lett. { 90} (2003) 140403.



\bibitem{Kasarev2}
K. Kasamatsu, M. Tsubota, and M. Ueda, 
Int. J. Mod. Phys. B {19} (2005) 1835.


\bibitem{Mueller}
E. J. Mueller and T. -L. Ho, 
Phys. Rev. Lett. {88} (2002) 180403.  
 

\bibitem{Barnett}
R. Barnett, G. Refael, M. A Porter, and H. P. Buchler, 
New J. Phys. { 10} (2008) 043030.

 
 

\bibitem{Wei}
R. Wei and E. Mueller,
Phys. Rev. A {84} (2011) 063611.

 

\bibitem{Kuopanportti}
P. Kuopanportti, J. A. M. Huhtam\"{a}ki, and M. M\"{o}tt\"{o}nen, 
Phys. Rev. A {85} (2012) 043613.




\bibitem{Kasashet}
K. Kasamatsu and M. Tsubota, 
Phys. Rev. A {79} (2009) 023606;

K. Kasamatsu, M. Tsubota, and M. Ueda, 
Phys. Rev. Lett. {91} (2003) 150406. 

 




\bibitem{phse}See, for example, 
V.P. Mineev,  Zh. Eksp. Teor. Fiz. 67 (1974) 263; [Eng. Transla. V.P. Mineev, Sov. Phys.-JETP 40 (1975) 132];

E. Timmermans, Phys. Rev.  Lett. { 81} (1998) 5718;

K. L. Lee,
N. B. Jorgensen,
I-K. Liu,
L. Wacker,
J. J. Arlt,
and N. P. Proukakis, Phys. Rev.  A { 94} (2016) 013602.



  \bibitem{quartic}A. L. Fetter, B. Jackson,
and S. Stringari, Phys. Rev.  A { 71} (2005) 013605.



 \bibitem{referee}H. Sakaguchi and B. A. Malomed,
Phys. Rev. A 78 (2008) 063606. 



\bibitem{ming}  G. Lamporesi, J. Catani, G. Barontini, Y. Nishida, M.
Inguscio, and F. Minardi, Phys. Rev. Lett. 104 (2010) 153202.




\bibitem{cs}T. Weber, J. Herbig, M. Mark, H.-C. N\"agerl, and  R. Grimm, Science 299 (2003) 232.


\bibitem{imag} P. Muruganandam and S. K. Adhikari, Comput. Phys.
Commun. { 180} (2009) 1888. 


\bibitem{ll1960}L. D. Landau,  and E. M. Lifshitz, 
Mechanics
(Pergamon Press, Oxford, 1960), section 39.


 
\bibitem{luca}L. Salasnich, A.  Parola, and L. Reatto,  
Phys. Rev.  A { 65} (2002) 043614.


   
\bibitem{CPC}  
D. Vudragovi\'c, I. Vidanovi\'c,
A. Bala\v z, P. Muruganandam, and S. K. Adhikari, Comput.
Phys. Commun. { 183} (2012) 2021;

L. E. Young-S., P. Muruganandam, S. K.  Adhikari, V. Lon\v car, D. Vudragovi\'c, and A.  Bala\v z, 
 Comput. Phys. Commun. { 220} (2017) 503;

V. Lon\v car, A.  Bala\v z, A. Boojevi\'c, S. \v Skrbi\'c, P. Muruganandam,  and S. K.  Adhikari, 
 Comput. Phys. Commun. { 200} (2016) 406;

L. E. Young-S., D. Vudragovi\'c, P. Muruganandam, S. K.  Adhikari, and  A.  Bala\v z, 
 Comput. Phys. Commun. { 204} (2016) 209;

B. Satari\'c, V. Slavnic, A. Beli\'c, A.  Bala\v z, P. Muruganandam,  and S. K.  Adhikari, 
 Comput. Phys. Commun. { 200} (2016) 411.
 
\bibitem{PS}P. Muruganandam and S. K. Adhikari, 
J. Phys. B { 36} (2003) 2501.


 
\bibitem{cpckk} R. Kishor Kumar, V. Lon\v car, A. Bala\v z, P. Muruganandam, and  S. K. Adhikari, unpublished (2018).
  

 

\bibitem{fesh}
S. Inouye, M. R. Andrews, J. Stenger, H.-J. Miesner, D. M. Stamper-Kurn, and W. Ketterle, Nature  { 392}  (1998) 151.






\end{thebibliography}
\end{document}